# Extended Depth of Field for High Resolution Scanning Transmission Electron Microscopy

Robert Hovden,[1] Huolin L. Xin,[2] and David A. Muller[1,3]

[1] *School of Applied and Engineering Physics, Cornell University, Ithaca, NY 14853*

[2] *Department of Physics, Cornell University, Ithaca, NY 14853*

[3] *Kavli Institute at Cornell for Nanoscale Science, Ithaca, NY 14853*

**Running Head**: Extended Depth of Field for STEM

**Corresponding Author**

Robert Hovden

212 Clark Hall, Cornell University, Ithaca, NY 14853

Tel: 607 255-0654

Fax: 607 255-7658

E-mail: rmh244@cornell.edu

**Key Words:** depth of field, aberration corrected, STEM, depth sectioning, deconvolution

**Abstract:** Aberration-corrected scanning transmission electron microscopes (STEM) provide sub-angstrom lateral resolution; however, the large convergence angle greatly reduces the depth of field. For microscopes with a small depth of field, information outside of the focal plane quickly becomes blurred and less defined. It may not be possible to image some samples entirely in focus. Extended depth-of-field techniques, however, allow a single image, with all areas in-focus, to be extracted from a series of images focused at a range of depths. In recent years, a variety of algorithmic approaches have been employed for bright field optical microscopy. Here, we demonstrate that some established optical microscopy methods can also be applied to extend the ~6 nm depth of focus of a 100 kV 5$^{th}$-order aberration-corrected STEM ($\alpha_{max}$ = 33 mrad) to image Pt-Co nanoparticles on a thick vulcanized carbon support. These techniques allow us to automatically obtain a single image with all the particles in focus as well as a complimentary topography map.

# INTRODUCTION

The recent introduction of aberration correctors has enabled scanning transmission electron microscope (STEM) imaging and analysis at sub-angstrom dimensions (Batson, 2006; Batson, et al., 2002; Erni, et al., 2009; Nellist, et al., 2004). In addition to the improved resolution, the large convergence angle greatly reduces the depth of field—the distance along the optical axis for which the sample is in focus (Born & Wolf, 1999). For microscopes with a small depth of field, information outside of the focal plane quickly becomes blurred and less defined. However, objects in focus are rendered sharper in the resulting image. The tradeoff between depth of field and lateral resolution can be problematic for biological samples with a large specimen thickness, especially for tomography (Hyun, et al., 2008); for determining the size distribution of catalysts nanoparticles on electrode supports; or high resolution electron microscopy where the depth of field becomes extremely small (Behan, et al., 2009; Xin, et al., 2008; Xin & Muller, 2009). The lateral resolution improves inversely proportional to the semi-angle of convergence, $\alpha_{max}$, while the depth of field has a more rapidly-diminishing inverse squared relationship. When working with aberration-corrected STEM with 0.70 Å resolution, the depth of focus is prohibitively small, ~6 nm (for a 100keV STEM). As a result, only parts of a typical TEM specimen (20-50 nm thick) will be in focus at the maximum resolution. At the same time, the numerical apertures are still not sufficient to allow reliable three-dimensional optical sectioning, with features distorted along the vertical axis by elongation factors of 30 – 50 (Behan, et al., 2009; Xin & Muller, 2009).

This would be equivalent to attempting tilt-series tomography with a ±2° tilt range (Intaraprasonk, et al., 2008; Xin & Muller, 2010).

We can extend the depth of field by moving the focal plane along the optical axis and acquiring images at each step, obtaining a $z$-stack. Ideally, to avoid under sampling, the distance between steps should be on the order of, or smaller than, the depth of focus of the point-spread function. The challenge then becomes extracting the in-focus information out of each image in the stack. That is, we would like to combine or merge the image stack into a single two-dimensional image with an apparent extended depth of field (Figure 1).

In this paper we analyze a typical fuel cell electrode structure - Pt-Co nano-particles attached to the outside of a three-dimensional aggregate of a vulcanized carbon support. This was imaged using a 100 kV $5^{th}$-order aberration-corrected STEM that has a depth of focus around ~6 nm. With the nano-particles roughly ~5 nm in diameter and well distributed over ~100 nm diameter carbon-black support particles, we cannot acquire a single image with all the nanoparticles simultaneously in focus for these operating conditions (Figure 2a, b). Through extended depth of field methods, a single image with all particles in focus allows large scale and meaningful counting statistics to be obtained for quantities such as particle size and spacing distributions, a process that would be far more time consuming if the full three-dimensional data set had to be analyzed.

# MATERIALS AND METHODS

Extended depth of field techniques used in light microscopy can be applied to annular dark-field STEM (ADF-STEM) fairly readily. There are three basic approaches to merging an image stack and obtaining an extended depth of field: A point-process basis, an area-process basis, and those that utilize the frequency space of the image (Forster, et al., 2004; Valdecasas, et al., 2001). All these approaches provide a z-height selection rule for each x-y location in the image stack. For transparent structures with multiple objects viewed in projection, the z-height function may no longer be singled valued, and methods that make that assumption could miss or misinterpret features. For this reason, we also consider deconvolution-based approaches.

*Point Processes Basis*

In the point process a single pixel at one location (transverse coordinates, x-y) is compared to others along $z$ in the image stack at the same ($x$, $y$) location. A selection rule, such as a gradient search, identifies a maximum intensity. Under a simple deconvolution model, this maximum intensity represents the z-height at which the slice is in focus. A simple maximum intensity and average intensity algorithm was used in the study for comparison.

*Area-Process Basis*

An area-based approach uses a neighborhood of pixels surrounding a location to determine the z-height at which the location is in focus. A classical example takes the variance of intensities in the neighborhood of each location: it is assumed that the z-height with the largest variance in intensity is in focus. It is particularly appealing because of its computational simplicity and relative effectiveness. There is a tradeoff in choosing a windows size for calculating the variance. A larger window size can provide more variance information but at a loss of locality. We used the variance over a 5x5 pixel window as a selection rule, although other sizes can also be used.

*Utilizing Frequency Space*

The frequency or wavelet approach has been successfully implemented by many groups for optical microscopy (Forster, et al., 2004; Unser & Aldroubi, 1996; Valdecasas, et al., 2001). It analyzes the frequency components of the image at various locations by using processes such as a wavelet or windowed Fourier transform. In-focus regions are assumed to have sharp details and contain high frequency information. Thus, regions with high frequency components are considered in focus and a z-height is selected from the stack.

The windowed Fourier transform uses a portion, or windowed region, of the sample to transform into frequency space. The limitation to this technique is the fixed window size which, when small, limits the frequency components or, when large, limits locality of the measurement. This is the underlying principle behind Heisenberg's uncertainty principle. As an alternative, we utilize a complex wavelet transform (Forster, et al., 2004; Unser & Aldroubi, 1996) . In the wavelet-based method, the image is transformed into a discrete basis of wavelets with frequency characteristics similar to the windowed Fourier transform (Mallat). Wavelets, however, have a fully-scalable modulated window. The magnitude of wavelet coefficients indicate the scale of detail in the image like that of the Fourier transform. The wavelet algorithm used here implements the complex Debauchies-6 wavelet basis (Forster, et al., 2004), which offers a smooth and continuous form. The complex wavelet transform of an image yields a coefficient for each scale, where large coefficients in the sub-bands correspond to high levels of detail. Each image in the stack is transformed and portions with largest coefficient magnitudes in the wavelet sub-bands are selected from in the *z*-stack. These selected components are then inverse wavelet transformed to yield the fused image with an extended depth of field.

*Deconvolution Methods*

Additionally, we examined the model-based deconvolution developed by F. Aguet, et al., where the image stack is assumed to be a convolution of the point spread function with a texture mapped to thin surface (Aguet, et al., 2008). The problem is thus to find a texture and topography map which, when convolved with the point spread function, yields the minimal intensity difference from the measured image stack. It becomes a least-squared optimization problem where the texture and topography maps are iteratively optimized. For reflected (episcopic) imaging like that of a Scanning Electron Microscope the thin surface approximation is trivially satisfied. For transmission (diascopic) microscopy, the approximation is only valid if the sample only appears in focus when focused on the sample surface. In the case of nano-particles, which are substantially small in comparison to the depth of focus, the non-overlapping particles can be approximated as a thin surface.

Three dimensional deconvolution models (Behan, et al., 2009; Nicolas, et al., 2004; Van den Broek, et al., 2010) do not assume a single in-focus plane, but rather attempt to reconstruct the sample's structure by deconvolving a point-spread function from an experimental image *z*-stack. A popular and common method for such a process is the Richardson-Lucy iterative algorithm (Lucy, 1974; Richards.Wh, 1972). It computes the maximum likelihood that an object, when convolved with a point spread function, will result in the original *z*-stack data—assuming Poisson noise statistics (Nicolas, et al., 2004). It has been shown that there are limitations to three-dimensional reconstructions from deconvolution (Xin & Muller, 2009) due to the large missing cone of information along the $k_z$ axis of the contrast transfer function. However, the $k_z = 0$ plane in Fourier space, which corresponds to the projected image, is complete with no missing information and is only attenuated by the instrument's contrast transfer function. This is the information needed to recover the extended depth of field. After deconvolving, the projected image is created by the average intensity along the axis of projection at each pixel location.

To test all of these algorithms, a sample of disperse Pt-Co nano-particles was used. The sample was chosen to have a relatively minimal overlap of particles along the axis of projection as to avoid the difficulty of two in-focus heights at the same x-y position. The Pt-Co nano-particles lie on a three-dimensional carbon black support that allows the particles to sit on a variety of focal planes within a region. The sample was prepared by micro-pipeting onto a lacy carbon grid before insertion into the microscope. Annular dark-field scanning tunneling electron microscope (ADF-STEM) images were acquired using a 100kV 5$^{th}$-order aberration-corrected STEM. The aberration-corrected machine's relatively large convergence angle of 33mrad provided a limited depth of focus around 5.8 nm. Acquiring a through-focal series of the region of interest was automated using an in-house DigitalMicrograph script that incrementally changes defocus by adjusting the high-tension voltage and records an image at each focus step. A series of 31 images with a 30nm defocus step were acquired and then aligned using cross correlation. Additionally, a second defocus series was taken on a single gold nano-particle (101 images with 4nm defocus step) with the in-focus images containing atomic detail. The gold nano-particle has appeared in (Xin & Muller, 2009).

# RESULTS

Extended depth of field algorithms can provide striking qualitative results with disperse nano-particles. Figures 2a,b show the drastic disparity of in-focus information at two different defocus values. When particles of one region are in focus, an adjacent region is dramatically blurred – and vice versa. A topological map of the particles in Figure 2c, generated from the wavelet approach discussed below, shows that the particles in the left and right regions lie at very different z-heights. This corresponds well with the through-focal data.

The extended depth-of-field algorithms presented previously were each applied to the through-focal series data in order to reconstruct a single in-focus image (Figure 2d-i). A simple point based, maximum-intensity approach shows the in-focus information of the high-intensity particles (Figure 2d), however this approach also picks up the out-of-focus information in the regions immediately surrounding particles. As a result, the particle edges can blend with the surrounding area, producing smaller than physical particle sizes. Another computationally simple approach—the variance method (Figure 2f)—accurately displayed the Pt-Co particles yet many artifacts were present elsewhere. Ringing artifacts around the particles are a result of the higher variance of the out-of-focus information's shot noise. Also, in the black carbon support and the central region of larger particles the low variance makes it difficult for the variance method to identify the in-focus information, often creating a topologically-noisy selection.

The wavelet approach produced better results (Figure 2e). All particles appear clearly in focus with their edges clearly defined. Additionally, the wavelet reconstruction performed well on a single nano-particle $z$-stack containing atomic resolution images. The high-frequency atomic information is preserved with the wavelet reconstruction (Figure 3c.), however there is a slight increase of background level when compared to the in-focus image of the support (Figure 3a,c,d,).

In the presence of noise, the wavelet algorithm picks up some of the high-frequency information of a particle's blur, as seen in the surrounding particle region(s) of Figure 2e. This also accounts for the splotchy or speckled appearance in the carbon support and vacuum regions. However, it seems that this speckling artifact is easily identifiable when interpreting the extended depth of field image and does not significantly detract from the reconstructed in-focus image. The wavelet algorithm was also able to maintain a high level of atomic detail in nano-particles (Figure 3c). Additionally, the wavelet method provides accurate topological identification of particles. By applying a threshold to the reconstructed extended depth of field image one can create an image mask and isolate the particle z-height information in the topology map (Figure 2c). The wavelet technique proves to be particularly appealing because it can produce high-quality results with no knowledge of the microscope's point spread function.

The Richardson-Lucy deconvolution approach was fairly successful at lower iterations (Figure 2h with 20 iterations). For higher iterations, edges become more defined at the cost of increased noise and reduction of contrast – as expected (Nicolas, et al., 2004). Ringing around particles from the deconvolution is present but the out-focus glow around particles is greatly reduced from that of the point-based maximum and average intensity approaches (Figure 2d and g) and particle edges are enhanced. The edges of the carbon support are preserved and the low variance regions appear homogenous. The atomic detail of the gold nano-particle was not well preserved in the reconstructed extended depth of field image (Figure 3b). Although Richardson-Lucy requires knowledge of the point spread function, it has the advantage of utilizing information in three dimensions. Unlike the other algorithms, no assumption is made that in-focus information occurs on a single plane. There are often two or more planes with in-focus information, as would be the case when two particles overlap along the optical axis. Figure 2(e,h) highlights two overlapping particles in which the overlapped intensity is only present in the deconvolution approach. For samples with higher particle density, a deconvolution approach may be the only acceptable extended depth of field technique.

The model-based deconvolution (Figure 2i), worked well when there were large regions with small topological variance. Artifacts were minimal, and the slowly varying and continuous topology prevents a splotchy appearance in the carbon support. Unfortunately, it has difficulty when there are overlapping particles lying on vastly different focal planes and therefore some particles appear out of focus or are missing entirely.

# DISCUSSION

Although a variety of extended depth of field approaches provide useful information, there is an opportunity for such techniques to be further improved or combined to better suit the needs of an electron microscopist. The primary difference of transmission electron microscopy from episcopic optical imaging is that, at a given x-y location, information can be held at two or more z-heights in an image stack. The techniques that have been presented, with the exception of Richardson-Lucy deconvolution, could be improved by allowing two or more in-focus planes to be included and averaged into the final fused extended depth of field image. Additionally, by adding heuristic rules, it may be possible to identify the homogeneous low noise vacuum regions present in a stack. Regions with noise-like topological variance could be punished slightly in favor of maintaining a smoother more continuous topology. It should also be noted that there are other techniques and filters which were not tested in this paper and we encourage the reader to explore (Burt, et al., 1993; Sobel & Feldman, 1968).

# CONCLUSIONS

Extended-depth-of-field techniques have been demonstrated as promising tools for aberration-corrected STEM. The small depth of field that accompanies high-resolution aberration corrected tools can be extended by taking a through-focal series and merging the in-focus information from the image stack. The nano-particles used in this paper represent a sample type that typically suffers from an electron microscope's limited depth of field. We have presented a variety of approaches - a popular subset of techniques found in optical microscopy - to create an in-focus 2D STEM image of nano-particles from a three dimensional through-focal series. The extended depth of field algorithms provided qualitative results (Figure 2e,h), where all particles appear in focus. Particular success was found for the complex-wavelet algorithm as well as a Richardson-Lucy deconvolution. However, it is important that the microscopist be aware of possible artifacts, in particular, the limited functionality of a single z-selection reconstruction, where only one plane of information at a given x-y position provides the in-focus information. Richardson-Lucy deconvolution is not limited to a single plane, and may therefore be more useful when there is a large number of overlapping particles. In summary, we have demonstrated that with the complex-wavelet algorithm or a Richardson-Lucy deconvolution, one can extend the effective depth of field when imaging a sparse distribution of nano-particles on a low-Z support.

## ACKNOWLEDGEMENTS

We gratefully acknowledge Julia A. Mundy for providing insight and careful revision to the manuscript. This work is supported by Semiconductor Research Corporation, the Cornell Center for Materials Research, an NSF MRSEC (DMR# 0520404) and the Energy Materials Center at Cornell, an Energy Frontier Research Center funded by the U.S. Department of Energy, Office of Basic Energy Sciences (DE-SC0001086).

# FIGURE CAPTIONS

Figure 1: Simulated particles at different heights (0, 15, 30 nm) are imaged over a range of defocuses (-4 to 34 nm) by a 100keV electron probe ($\alpha_{max}$ = 33 mrad). Due to the microscope's limited depth of field (*shown left*), particles go in and out of focus. By merging the in-focus information from each image in the stack, we can effectively extend the depth of focus (*shown right*).

Figure 2: Pt-Co nano-particles imaged by an aberration-corrected Nion UltraSTEM (100keV, 33mrad). a) and b) shows two regions with different focal planes from the original focal series. c) shows the particle topography obtained from the wavelet method. The depth of field was extended using: d) point-based max intensity, e) wavelet transforms, f) variance method, g) an averaged stack, h) a summed Richardson-Lucy deconvolution stack, and i) a model-based deconvolution. Arrows highlights the difference in how the wavelet and deconvolution approaches handle two overlapping particles as more clearly seen in the enlarged inset images (e,h).

Figure 3: A through-focal-series z-stack of a gold nano-particle acquired with atomic resolution. The in-focus image (a), the summed RL deconvolution after 20 iterations (b), and the reconstructed complex wavelet extended depth of field images (c) are shown. The inset images in (a) and (b) show cross sections through the z-stack before and after deconvolution. The high frequency atomic information is preserved well using the wavelet approach, but there is a slight increase of background level of the support - as seen in the comparative line profile along the vertical direction averaged across ten pixels (d).

# FIGURES

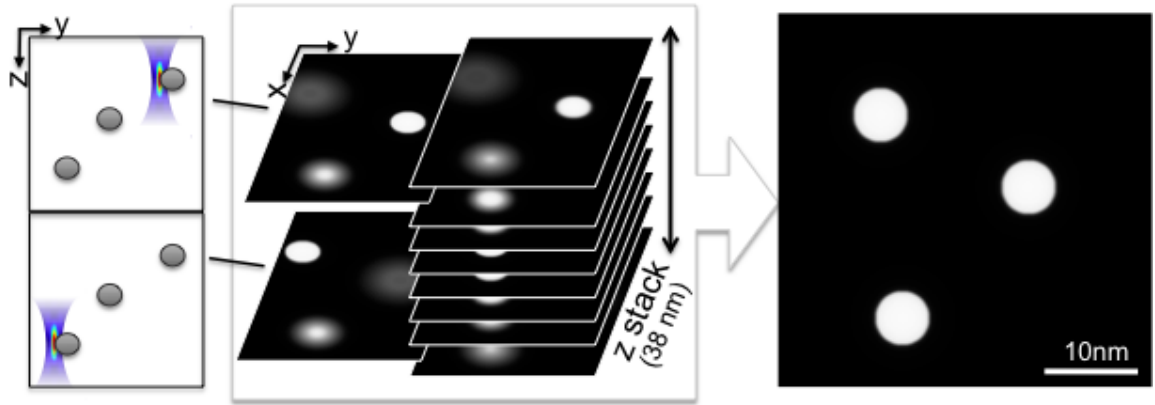

Figure 1: Simulated particles at different heights (0, 15, 30 nm) are imaged over a range of defocuses (-4 to 34 nm) by a 100keV electron probe ($\alpha_{max}$ = 33 mrad). Due to the microscope's limited depth of field (*shown left*), particles go in and out of focus. By merging the in-focus information from each image in the stack, we can effectively extend the depth of focus (*shown right*).

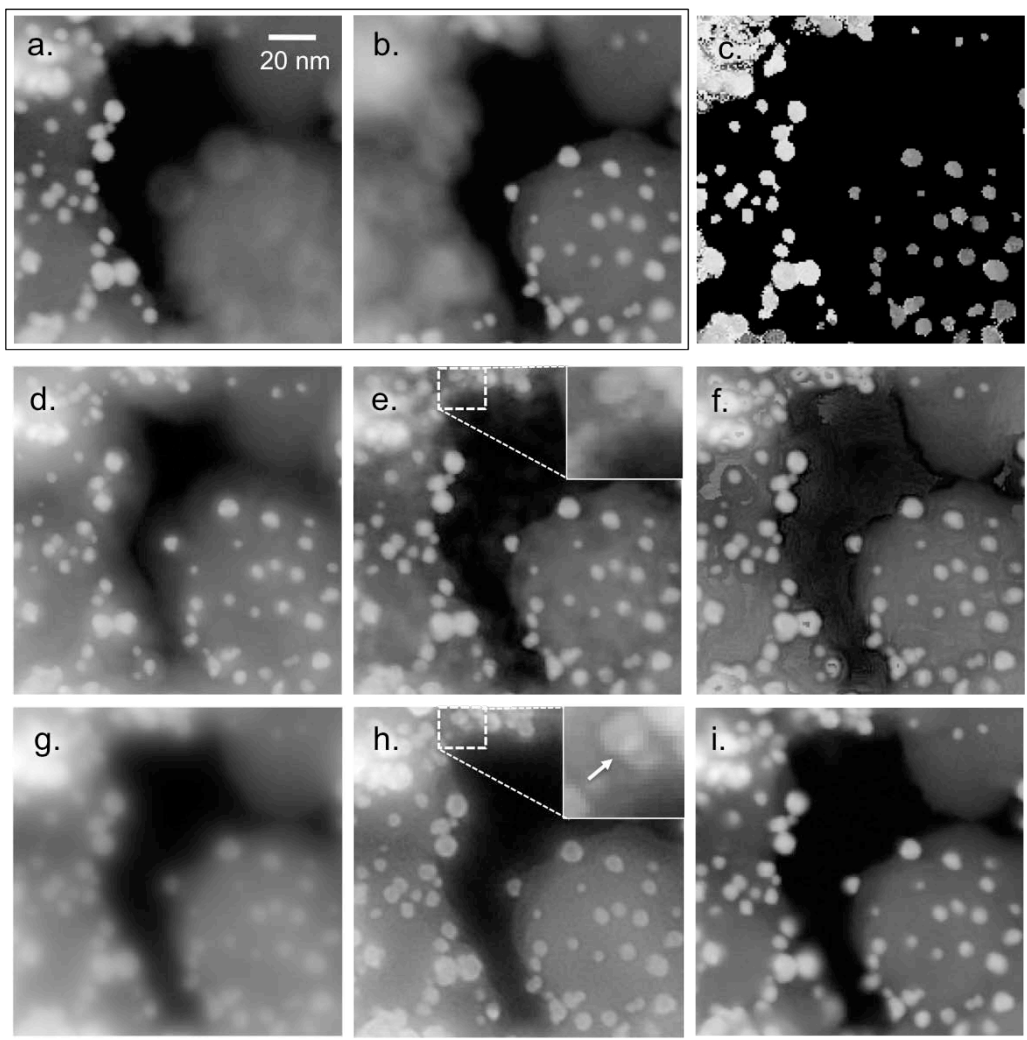

Figure **2**: Pt-Co nano-particles imaged by an aberration-corrected Nion UltraSTEM (100keV, 33mrad). a) and b) shows two regions with different focal planes from the original focal series. c) shows the particle topography obtained from the wavelet method. The depth of field was extended using: d) point-based max intensity, e) wavelet transforms, f) variance method, g) an averaged stack, h) a summed Richardson-Lucy deconvolution stack, and i) a model-based deconvolution. Arrows highlights the difference in how the wavelet and deconvolution approaches handle two overlapping particles as more clearly seen in the enlarged inset images (e,h).

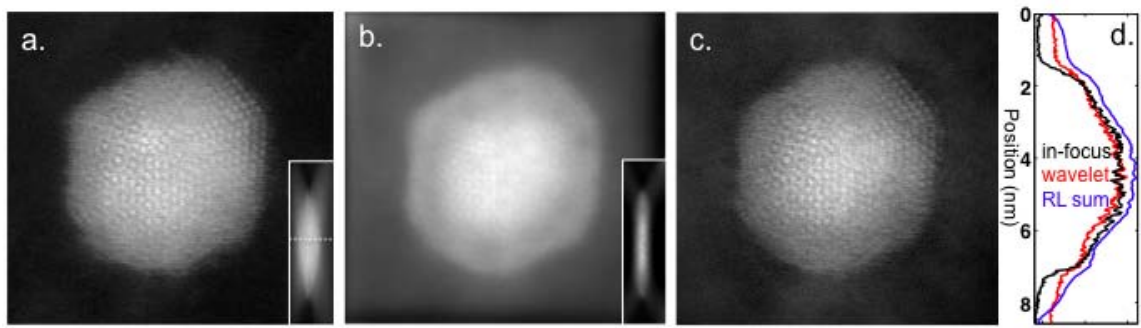

Figure **3**: A through-focal-series z-stack of a gold nano-particle acquired with atomic resolution. The in-focus image (a), the summed RL deconvolution after 20 iterations (b), and the reconstructed complex wavelet extended depth of field images (c) are shown. The inset images in (a) and (b) show cross sections through the z-stack before and after deconvolution. The high frequency atomic information is preserved well using the wavelet approach, but there is a slight increase of background level of the support - as seen in the comparative line profile along the vertical direction averaged across ten pixels (d).